\documentclass [twocolumn,aps,prb]{revtex4}
\usepackage{graphicx,color,amssymb,amsmath,multirow,array}
\begin{document}
\title{Probing Interband Coulomb Interactions in Semiconductor Nanocrystals with 2D Double-Quantum Coherence Spectroscopy }
\author{Kirill A. Velizhanin}
\author{Andrei Piryatinski}\email{apiryat@lanl.gov}

\address{Center for Nonlinear Studies (CNLS),
 Theoretical Division, Los Alamos National Laboratory, Los Alamos, NM 87545}

\date{\today}

\begin{abstract}
Using previously developed exciton scattering model accounting for the interband, i.e., exciton-biexciton, Coulomb interactions in semiconductor nanocrystals (NCs),	 we derive a closed set of equations for 2D double-quantum coherence signal. The signal depends on the Liouville space pathways which include both the interband scattering processes and the inter- and intraband optical transitions. These processes correspond to the formation of different cross-peaks in the 2D spectra. We further report on our numerical calculations of the 2D signal using reduced level scheme parameterized for PbSe NCs. Two different NC excitation regimes considered and unique spectroscopic features associated with the interband Coulomb interactions are identified.
\end{abstract}

\maketitle

\section{Introduction}
\label{intro}

Coulomb interactions between carriers in semiconductor materials including nanocrystals (NCs) play central role in photoexcited dynamics.\cite{haug09,haug08} Introducing conduction and valence bands within mean field (e.g., effective mass) approximation, these interactions can be partitioned into two groups.\cite{axt98} First group combines the interactions which conserve the number of electrons/holes, and can be associated  with the diagonal terms in the electron-hole interaction Hamiltonian. In other words, these interactions are acting within exciton bands of different multiplicity, and therefore, will be referred as the {\em intraband} Coulomb interactions.  The intraband interactions determine the binding energies of excitons, trions, and multi-excitons.\cite{efros89,banyai88,banyai89,takagahara89,hu90} Spectroscopically, these quantities can be measured as energy shifts in transient-absorption and fluorescent techniques. One of the important areas where such interactions are widely exploited is wave-function engineering in semiconductor heterostructures \cite{butov02,piryatinski07a} with applications to the problem of the lasing in single-exciton regime.\cite{klimov07}     

Second group includes off-diagonal electron-hole interaction terms in the Hamiltonian. These interactions lead to the valence-conduction band transitions which change the number of electrons and holes conserving the total charge. In other words, the off-diagonal Coulomb interactions are responsible for the photoexcited dynamics which involves transitions between exciton bands of different multiplicity. Therefore, we refer to such interactions as the {\it interband} Coulomb interactions. Important examples involving the interband transitions are non-radiative Auger recombination and impact ionization processes.\cite{landsberg03}

Lately, carrier multiplication (CM) in semiconductor NCs has received significant attention.\cite{schaller04,ellingson05,nair08,mcguire08,pijpers09} Also referred as the multi-exciton generation, this is a process of more than one electron-hole pair generation per absorbed photon which has great potential for applications in photovoltaic, photochemical, and energy storage devices.\cite{kolodinski93,nozik02,hanna06,klimov06,beard08a,nozik08,luther08} A number of models for CM dynamics based on the interband Coulomb interactions which go beyond simple impact ionization model have been proposed recently.\cite{schaller05,rupasov07,shabaev06,witzel10,piryatinski10} Direct experimental technique for measuring interband Coulomb interactions in NCs should greatly advance our understanding of CM by shedding new light on the multi-exciton photogeneration dynamics and providing verification of the proposed theoretical models. 

Since CM takes place in the region of high exciton and biexciton DOS, key features associated with a variety of the inter- and intraband interactions are spectrally overlapped and cannot be resolved using conventional techniques, e.g., transient absorption and time-resolved fluorescence spectroscopies. In this case, the resolution can be tremendously enhanced by using coherent ultrafast nonlinear techniques in which various interaction pathways are distinguished through the spatial phase-matching conditions.\cite{mukamel_book} Further representing the coherent signal as a 2D Fourier transform polarization with respect to various pulse delay times, one can gain direct insight into the hidden interactions by looking at the position, line-shape, intensity and phase of the associated cross-peaks.\cite{mukamel00,mukamel04,li06,mukamel07,oszwaldowski08} The latter techniques are referred as 2D correlation spectroscopies and represent ultrafst optical counterpart of 2D NMR.\cite{ernst90}  

Among 2D correlation spectroscopies, the double-quantum coherence technique proposed by Mukamel and co-workers is considered to be most sensitive to the many-body correlations.\cite{mukamel07a} This technique has been applied to study the intrerband Coulomb correlations leading to biexciton formation in organic molecular systems,\cite{abramavicius08,kim09} and semiconductor quantum wells\cite{yang08,yang08a,stone09,stone09a,karaiskaj10}. The application of this methodology to probe interband Coulomb interactions, requires additional theoretical study which is the focus of this paper. 

To model nonlinear optical response of coupled exciton-biexciton states, we adopt the exciton scattering model previously developed by us to provide a unified treatment of CM dynamics in semiconductor NCs.\cite{piryatinski10} To avoid difficulties associated with large number of resonances in the region of high-DOS manifold and to unambiguously identify 2D spectroscopic signatures of the interband interactions, we performed numerical calculations using reduced level scheme with the parameters for PbSe NCs.\cite{kang97} 

The paper is organized as follows: In Sec.~\ref{theory}, we discuss our general theoretical model for the nonlinear optical response of coupled exciton-biexciton states and spectroscopic features immediately following form the theory. In Sec.~\ref{numer}, the  numerical calculations of double-quantum coherence signal for PbSe NCs are discussed in details. Finally, we present our conclusions in Sec.~\ref{conc}.

\section{Nonlinear response from coupled exciton and biexciton states}
\label{theory}

Let us consider an ensemble of NCs, in which the carrier dynamics is restricted to the coupled exciton and biexciton manifolds described by the following projected Hamiltonian\cite{piryatinski10}
\begin{eqnarray}\label{Hs}
 \hat{H} &=& \hat{H}_0+\hat V_C.
\end{eqnarray}
Here, the first term,
\begin{eqnarray}\label{H0}
 \hat{H}_0&=& \sum_{a}|x_a\rangle\hbar\omega^x_a\langle x_a|
    + \sum_{k}|xx_k\rangle\hbar\omega^{xx}_k\langle xx_k|,
\end{eqnarray}
describes non-interacting exciton, $|x_a\rangle$, and biexciton, $|xx_k\rangle$, states characterized by the energies $\hbar\omega^x_a$ and $\hbar\omega^{xx}_k$, respectively. These energies already include corresponding binding energies associated with the {\em intraband} Coulomb interactions. The second term
\begin{eqnarray}\label{HW}
 \hat{V}_C &=& \sum_{a}\sum_{k}|x_a\rangle V^{x,xx}_{a,\;k}\langle xx_k|
 +h.c.,
\end{eqnarray}
represents the {\em interband} Coulomb interactions, $V^{x,xx}_{a,k}$, between the states. In this paper, we drop the vacuum, $|x_0\rangle$, to biexciton couplings, $V^{xx,x}_{k,0}$, considered in the CM theory,\cite{rupasov07,piryatinski10} since their contribution to the nonlinear optical response is negligible. Explicit representation for interaction matrix elements, $V^{x,xx}_{a,\;k}$, in terms of the single-particle couplings, and related matrix equations defining the exciton and biexciton states can be found in Ref~[\onlinecite{piryatinski10}].

Within the exciton scattering model, coupled exciton and biexciton state propagation is described by the Hilbert space Green function\cite{piryatinski10}
\begin{eqnarray}\label{G-mtrx}
 \hat G(t) = \left( \begin{array}{cc}
					\hat G^{x}(t)    & \hat G^{x,xx}(t)\\
					\hat G^{xx,x}(t) & \hat G^{xx}(t)  \\
	\end{array} \right),
\end{eqnarray}
whose diagonal blocks determine the intraband transitions renormalized by the even-order interband scattering events. Since, these propagators do not change the exciton state multiplicity, we will refer to the associated processes as {\em intraband propagation}. The off-diagonal components describe the odd-order interband scattering events changing the exciton multiplicity. As a result, associated  processes will be referred below as the {\em interband propagation}. A closed set of equations for the Green function based on the scattering matrix formalism is provided in Appendix~\ref{app-esm}.  

\begin{figure}[t]
\centering
\includegraphics[width=3.0in,clip]{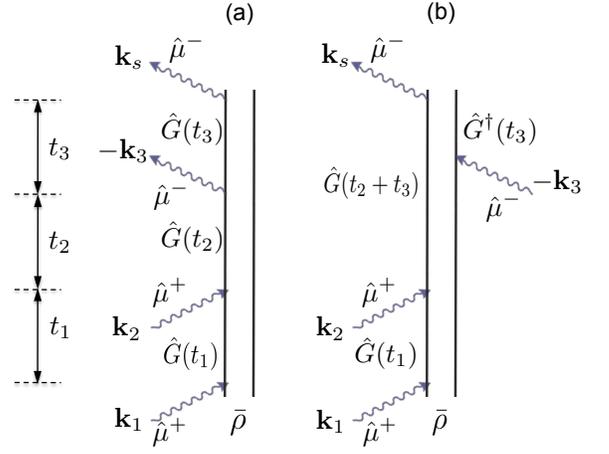}
\caption{Double-sided Feynman diagrams associated with Eq.~(\ref{Ptd}) for the 
		third-order nonlinear optical polarization characterized by the wave vector 
		${\bf k}_s={\bf k}_1+{\bf k}_2-{\bf k}_3$.
}
\label{fig-Fd}
\end{figure}

The interaction of coupled exciton and biexciton states with optical field is described by the extended Hamiltonian 
\begin{eqnarray}\label{Hopt}
 \hat{H}_{opt} = \hat{H} - \hat\mu E(t),
\end{eqnarray}
where $\hat{H}$ is the material part (Eqs.~(\ref{Hs})-(\ref{HW})). The second term in Eq.~(\ref{Hopt}) contains electric field 
\begin{eqnarray}\label{ptrain}
	E({\bf r},t)=\sum_{j=1}^4{\cal E}_j(t-\tau_j)e^{i{\bf k}_i{\bf\cdot r}-i\omega_0 t}+c.c..
\end{eqnarray}
which is a sequence of four linearly polarized non-overlapping ultrafast pulses. The first three pulses excite the third order nonlinear polarization and the fourth one is used to heterodyne the signal.\cite{mukamel_book} Each pulse is characterized by a wave vector ${\bf k}_i$ and central frequency $\omega_0$ which is set identical for the whole train. We assume impulsive excitation regime, and approximate the envelope functions as ${\cal E}_j(t-\tau_i)={\cal E}_0\delta(t-\tau_i)$, where $\delta(t-\tau_i)$ is the Dirac delta-function and ${\cal E}_0$ is effective amplitude. The latter without lack of generality is set to ${\cal E}_0=1$.   

The dipole moment operator projected on the optical field in Eq.~(\ref{Hopt}) has the following block-matrix form 
\begin{eqnarray}\label{mu-mtrx}
 \hat\mu = \left( \begin{array}{cc}
					\hat\mu^{x}    &\hat\mu^{x,xx}\\
					\hat\mu^{xx,x} & \hat\mu^{xx}\\
	\end{array} \right),
\end{eqnarray}
where the upper (lower) diagonal block describes the interband vacuum-exciton transitions and the intraband exciton (biexciton) transitions. The off-diagonal blocks describe interband exciton-biexciton transitions.    

Using the sum-over-states approach along with the rotating-wave approximation\cite{mukamel07}, we derived the following expression for the nonlinear polarization propagating along the ${\bf k}_s={\bf k}_1+{\bf k}_2-{\bf k}_3$ direction
\begin{eqnarray}\label{Ptd}
 P(t_3,t_2,t_1) &=& \left(Tr[\hat\mu^{-}\hat G(t_3)\hat\mu^{-} \hat G(t_2)\hat\mu^{+}\hat G(t_1)\hat\mu^{+}\bar\rho]
 \right.\\\nonumber&-&\left.
 Tr[\hat\mu^{-}\hat G^\dag(t_3)\hat\mu^{-}\hat G(t_3+t_2)\hat\mu^{+} \hat G(t_1)\hat\mu^{+}\bar\rho]\right)
 \\\nonumber &\times&
 e^{i\omega_0t_3+2i\omega_0t_2+i\omega_0t_1},
 \end{eqnarray}
containing introduced Hilbert space Green functions, $G(t)$ . In Eq.~(\ref{Ptd}), the dipole operator (Eq.~(\ref{mu-mtrx})) is partitioned into the sum of two terms, $\hat\mu=\hat\mu^{+}+\hat\mu^{-}$, describing optical transitions up and down, respectively.\cite{mukamel07} $\bar\rho=|x_0\rangle\langle x_0|$ is the equilibrium density operator for the ensemble of NCs in the vacuum state.\footnote{The exciton scattering models for CM\cite{piryatinski10} provides up to the second order corrections to this operator in terms of the interband Coulomb coupling. However, the contribution of these terms to the nonlinear response is negligible.} 

The first (second) term in Eq.~(\ref{Ptd}) are associated with the Liouville space pathway schematically depicted by first (second) double-sided Feynman diagram in Fig.~(\ref{fig-Fd}). These pathways have been studied before for exciton and biexciton manifolds, where the intraband dipole transitions and the interband propagations are forbidden.\cite{mukamel07,abramavicius08,yang08,yang08a,karaiskaj10} However, in NCs these processes should be accounted for and, as shown below, lead to new features in  double-quantum coherence signal, which is defined as 
\begin{eqnarray}\label{S2D}
 S(\Omega_3,\Omega_2) &=& \int_0^\infty dt_3\int_0^\infty dt_2 P(t_3,t_2,t_1=0)
 \\\nonumber &\times&
 			e^{i(\Omega_3-\omega_0)t_3+i(\Omega_2-2\omega_0)t_2}.
\end{eqnarray}
Note that for further convenience, the origin of $(\Omega_3,\Omega_2)$ 2D signal representation is shifted to $(\omega_0,2\omega_0)$, where $\omega_0$ is pulse central frequency.

\begin{figure}[t]
\centering
\includegraphics[width=3.2in,clip]{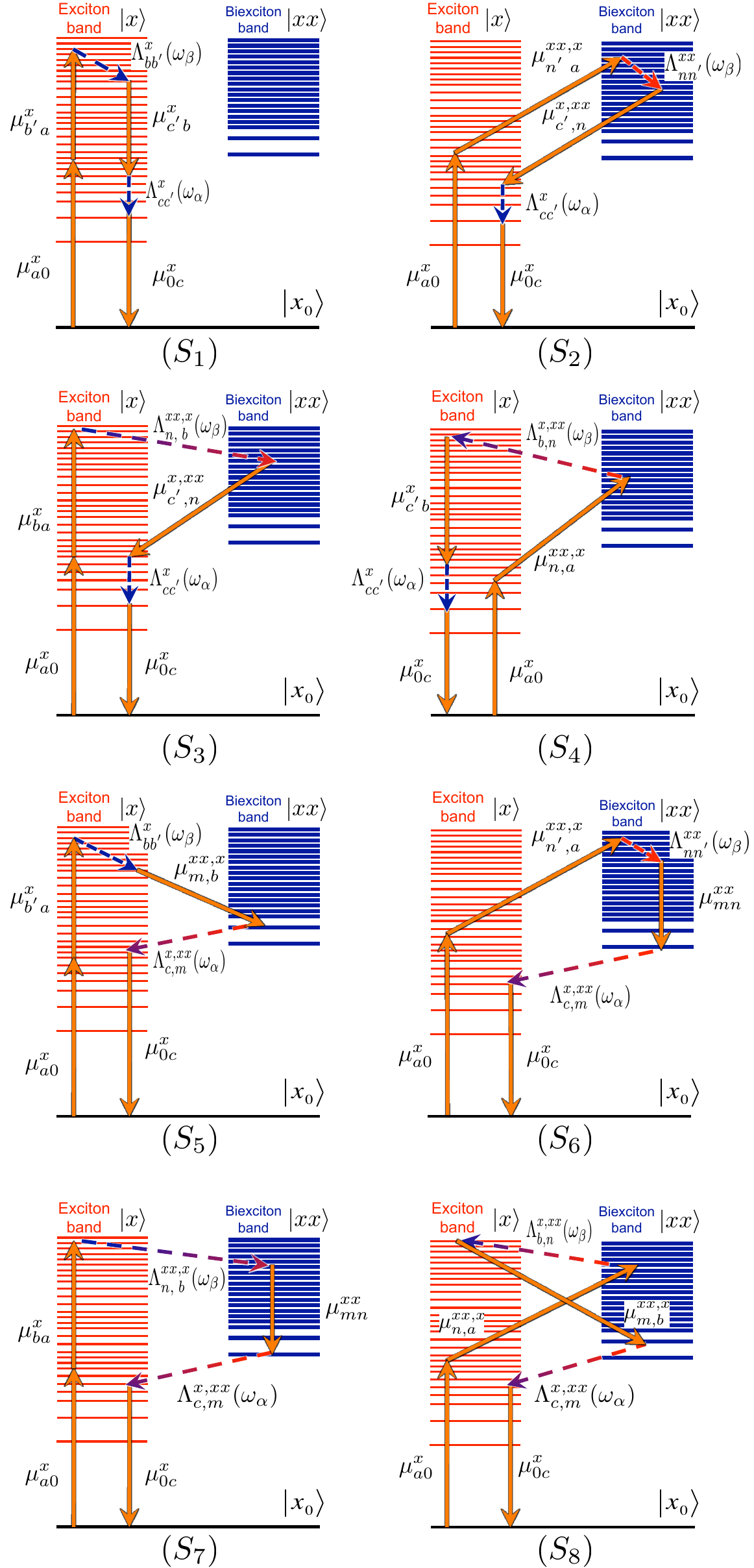}
\caption{Level diagrams representing optical transition and coherence propagation pathways for the components of double-quantum coherence signal given by Eqs.~(\ref{Sall}) and (\ref{Sj}). Solid arrows represent optically induced transitions while the dashed arrows show coherence propagations during pulse delay times. Notations for the dipole moment elements and transition amplitudes are the same as used in Table~\ref{tab-Sj}.} 
\label{fig-ld}
\end{figure}

To calculate double-quantum coherence signal according to Eqs.~(\ref{Ptd}) and (\ref{S2D}), one needs to find the Green function (Eq.~(\ref{G-mtrx})) by solving numerically  set of equations provided in Appendix~\ref{app-esm}. However, this general approach does not give insight into the cross-peak structure. Therefore, we apply a different approach using the following representation for the diagonal blocks of the Hilbert space Green function\cite{piryatinski10}   
\begin{eqnarray}
\label{GXL}
    \hat G^{x}(t)&=&\sum_{\xi}\sum_{ab}|x_a\rangle\Lambda^x_{ab}(\omega_{\xi})\langle x_b| e^{-i\tilde\omega_{\xi}t},
\\\label{GXXL}
    	\hat G^{xx}(t)&=&\sum_{\xi}\sum_{kl}|xx_k\rangle\Lambda^{xx}_{kl}(\omega_{\xi})\langle xx_l|e^{-i\tilde\omega_{\xi}t}.
\end{eqnarray}
Here, complex frequencies, $\tilde\omega_{\xi} = \omega_{\xi}-i\gamma_{\xi}$, are the poles of $\hat G(\omega)$, representing the eigenfrequencies of the Coulomb coupled exciton and biexciton states. Hereafter, we will refer to these states, $|\xi\rangle$, as the {\em quasiparticle} states. $\Lambda^x_{ab}(\omega_{\xi})={\rm res}\{G^x_{ab}(\tilde\omega_{\xi})\}$   ($\Lambda^{xx}_{kl}(\omega_{\xi})={\rm res}\{G^{xx}_{kl}(\tilde\omega_{\xi})\}$) are exciton (biexciton) Green function residues determining probability amplitudes for the transition between $|x_a\rangle$ and $|x_b\rangle$  ($|xx_k\rangle$ and $|xx_l\rangle$) states associated with the time evolution of a quasiparticle state $|\xi\rangle$.

\begin{table*}[t]
   \centering
   \begin{tabular}{m{1.5cm}|c|c|l|l} 
\hline\hline
Number of Interband Propagation Events	&\;\;$S_j$\;\;  &\;$q_j$\;&\hspace{1.2cm}$A_{q_j}(\omega_\alpha)$&\hspace{1.2cm}$B_{q_j}(\omega_\beta)$\\
\hline\vspace{0.25cm}
\multirow{2}{*}{\hspace{0.7cm}0}& $S_1$  & $b$ &\;\;$\sum_{cc^{'}}\mu^x_{0c}\Lambda^{x}_{cc^{'}}(\omega_\alpha)\mu^{x}_{c^{'}b}$\;\; &\;\;$\sum_{ab^{'}}\Lambda^x_{bb^{'}}(\omega_\beta)\mu^x_{b^{'}a}\mu^x_{a0}$\;\;\\[0.25cm]
	 &$S_2$  & $n$ &\;\;$\sum_{cc^{'}}\mu^x_{0c}\Lambda^{x}_{cc^{'}}(\omega_\alpha)\mu^{x,xx}_{c^{'},n}$\;\; &\;\;$\sum_{an^{'}}\Lambda^{xx}_{nn^{'}}(\omega_\beta)\mu^{xx,x}_{n^{'},a}\mu^x_{a0}$\;\;\\[0.25cm]
\hline\vspace{0.25cm}
\multirow{4}{*}{\hspace{0.7cm}1}& $S_3$   & $n$ &\;\;$\sum_{cc^{'}}\mu^x_{0c}\Lambda^{x}_{cc^{'}}(\omega_\alpha)\mu^{x,xx}_{c^{'},n}$\;\; &\;\;$\sum_{ab}\Lambda^{xx,x}_{n,\;b}(\omega_\beta)\mu^{x}_{ba}\mu^x_{a0}$\;\;\\[0.25cm]
 	& $S_4$   & $b$ &\;\;$\sum_{cc^{'}}\mu^x_{0c}\Lambda^{x}_{cc^{'}}(\omega_\alpha)\mu^{x}_{c^{'}b}$\;\; &\;\;$\sum_{an}\Lambda^{x,xx}_{b,\;n}(\omega_\beta)\mu^{xx,x}_{n,a}\mu^x_{a0}$\;\;\\[0.25cm]
 	& $S_5$   & $b$ &\;\;$\sum_{cm}\mu^x_{0c}\Lambda^{x,xx}_{c,m}(\omega_\alpha)\mu^{xx,x}_{m,b}$\;\; &\;\;$\sum_{ab^{'}}\Lambda^{x}_{bb^{'}}(\omega_\beta)\mu^{x}_{b^{'}a}\mu^x_{a0}$\;\;\\[0.25cm]
 	& $S_6$   & $n$ &\;\;$\sum_{cm}\mu^x_{0c}\Lambda^{x,xx}_{c,m}(\omega_\alpha)\mu^{xx}_{m,n}$\;\; &\;\;$\sum_{an^{'}}\Lambda^{xx}_{nn^{'}}(\omega_\beta)\mu^{xx,x}_{n^{'},a}\mu^x_{a0}$\;\;\\[0.25cm]
\hline\vspace{0.25cm}
\multirow{2}{*}{\hspace{0.7cm}2}	& $S_7$   & $n$ &\;\;$\sum_{cm}\mu^x_{0c}\Lambda^{x,xx}_{c,m}(\omega_\alpha)\mu^{xx}_{mn}$\;\; &\;\;$\sum_{ab}\Lambda^{xx,x}_{n,\;b}(\omega_\beta)\mu^{x}_{ba}\mu^x_{a0}$\;\;\\[0.25cm]
 	& $S_8$   & $b$ &\;\;$\sum_{cm}\mu^x_{0c}\Lambda^{x,xx}_{c,m}(\omega_\alpha)\mu^{xx,x}_{m,b}$\;\; &\;\;$\sum_{an}\Lambda^{x,xx}_{b,n}(\omega_\beta)\mu^{xx,x}_{\;n,\;a}\mu^x_{a0}$\;\;\\[0.25cm]
\hline\hline  
   \end{tabular}
   \caption{List of coefficients associated with the double-quantum coherence signal components given in Eq.~(\ref{Sj}).
   }
   \label{tab-Sj}
\end{table*}

Similar expansion for the Green function off-diagonal blocks reads
\begin{eqnarray}
\label{GXXXL}
    \hat G^{x,xx}(t)&=& \sum_{a,k}|x_a\rangle\Lambda^{x,xx}_{a,k}(\omega_{\xi})\langle xx_k|e^{-i\tilde\omega_{\xi}t},
\end{eqnarray}
where $\Lambda^{x,xx}_{a,k}(\omega_{\xi})={\rm res}\{G^{x,xx}_{a,k}(\tilde\omega_{\xi})\}$ are the transition probability amplitudes between exciton, $|x_a\rangle$, and biexciton, $|xx_k\rangle$, states associated with $|\xi\rangle$. Adopted representation (Eqs.~(\ref{GXL})--(\ref{GXXXL})) provides connection between the spectroscopic resonances determined by the quasiparticle frequencies and exciton/biexciton dynamics described by the associated transition amplitudes. 

Using Eqs.~(\ref{GXL})--(\ref{GXXXL}) along with Eqs.~(\ref{mu-mtrx})--(\ref{S2D}), one finds that double-quantum coherence signal can be represented as a sum of eight components
\begin{eqnarray}\label{Sall}
 S(\Omega_3,\Omega_2) =\sum_{j=1}^8 S_{j}(\Omega_3,\Omega_2).
\end{eqnarray}
Each component in Eq.~(\ref{Sall}) can be attributed to specific transition (Liouville space) pathway shown in level diagrams of Fig.~\ref{fig-ld}. Mathematically, signal components can be represented in the following compact form
\begin{eqnarray}\label{Sj}
 S_{j}(\Omega_3,\Omega_2) &=& \sum_{\alpha\beta}\sum_{q_j}
 \frac{B_{q_j}(\tilde\omega_\beta)}{\Omega_2-\tilde\omega_{\beta}}
 \\\nonumber&\times&
 \left(\frac{A_{q_j}(\tilde\omega_\alpha)}	{\Omega_3-\tilde\omega_{\alpha}}
 		-\frac{A^*_{q_j}(\tilde\omega_\alpha)}{\Omega_3-\tilde\omega_{\beta\alpha}}\right),
\end{eqnarray}
where $\tilde\omega_{\beta\alpha}=\tilde\omega_\beta-\tilde\omega_{\alpha}^*$. Expressions for the coefficients $A_{q_j}(\tilde\omega_\alpha)$ and $B_{q_j}(\tilde\omega_\beta)$ in terms of the dipole matrix elements and the transition amplitudes are listed in Table~\ref{tab-Sj}.

According to Fig.~\ref{fig-ld}, the signal components $S_1$ and $S_2$ depend on the transition pathways which comprise only the intraband propagation events following the dipole transition and no interband propagation. This shows up in Table~\ref{tab-Sj} as the dependence of the associated coefficients $A_{q_j}(\tilde\omega_\alpha)$ and $B_{q_j}(\tilde\omega_\beta)$ on intraband transition amplitudes $\Lambda^{x}$ and $\Lambda^{xx}$ only. The interband transition amplitude, $\Lambda^{x,xx}$, appears in  $B_{q_j}(\tilde\omega_\beta)$ ($A_{q_j}(\tilde\omega_\alpha)$) for the signal components $S_3$ and $S_4$ ($S_5$ and $S_6$) and reflects an interband propagation event in high, i.e, $\omega_\beta\sim 2\omega_0$, (low, i.e., $\omega_\alpha\sim \omega_0$) frequency region (Fig.~\ref{fig-ld}). Finally, the last two components $S_7$ and $S_8$ depend on the pathways which involve both high and low frequency interband propagation events (Fig.~\ref{fig-ld}).   

Using Eq.~(\ref{Sj}) and Table~\ref{tab-Sj}, a general analysis of the cross-peak structure can be performed. For this purpose, we initially assume that the interband Coulomb interaction is turned off. In this case $\Lambda^{x,xx}=0$, and thus, all signal components vanish except $S_1$ and $S_2$.  Further taking into account that $\Lambda^{x}_{ab}(\omega_\xi)=\delta_{ab}\Lambda^{x}_{aa}(\omega^{x}_a)$ ($\Lambda^{xx}_{kl}(\omega_\xi)=\delta_{kl}\Lambda^{xx}_{ll}(\omega^{xx}_k)$), entering expressions for $S_1$ and $S_2$, become dependent on the {\em unperturbed} exciton (biexciton) frequencies, the following two groups of $(\Omega_3,\Omega_2)$-cross-peaks can be identified:  $(\omega^{x}_b,\omega^{x}_c)$ and $(\omega^{x}_b,\omega^{x}_{cb})$ due to $S_1(\Omega_3,\Omega_2)$, and  $(\omega^{xx}_n,\omega^{x}_c)$ and $(\omega^{xx}_n,\omega^{x,xx}_{n,c})$ due to $S_2(\Omega_3,\Omega_2)$. Note, that as a result of the intraband exciton transitions (Fig.~\ref{fig-ld}), the cross-peaks associated with $S_1$ depend on the exciton frequency only.  

Turning on the interband Coulomb interactions leads to the mixing of the exciton and biexciton states and splitting of $\omega^{x}_a$ and $\omega^{xx}_k$ frequencies into quasiparticle frequencies $\omega_\xi$. The splitting lead to the appearance of new cross-peaks in all components of the signal. Further analysis of new cross-peaks requires their identification using simplified level structure. Before proceeding with this analysis, we notice that in the important weak Coulomb limit of the exciton scattering model multiple scattering events between coupled states reduce to a single (Born) scattering event.\cite{piryatinski10} Expressions for double-quantum coherence signal in this limit are derived in Appendix~\ref{app-wcl}.

\section{Numerical Calculations and Discussion}
\label{numer}

\begin{figure}[t]
\centering
\includegraphics[width=3.4in,clip]{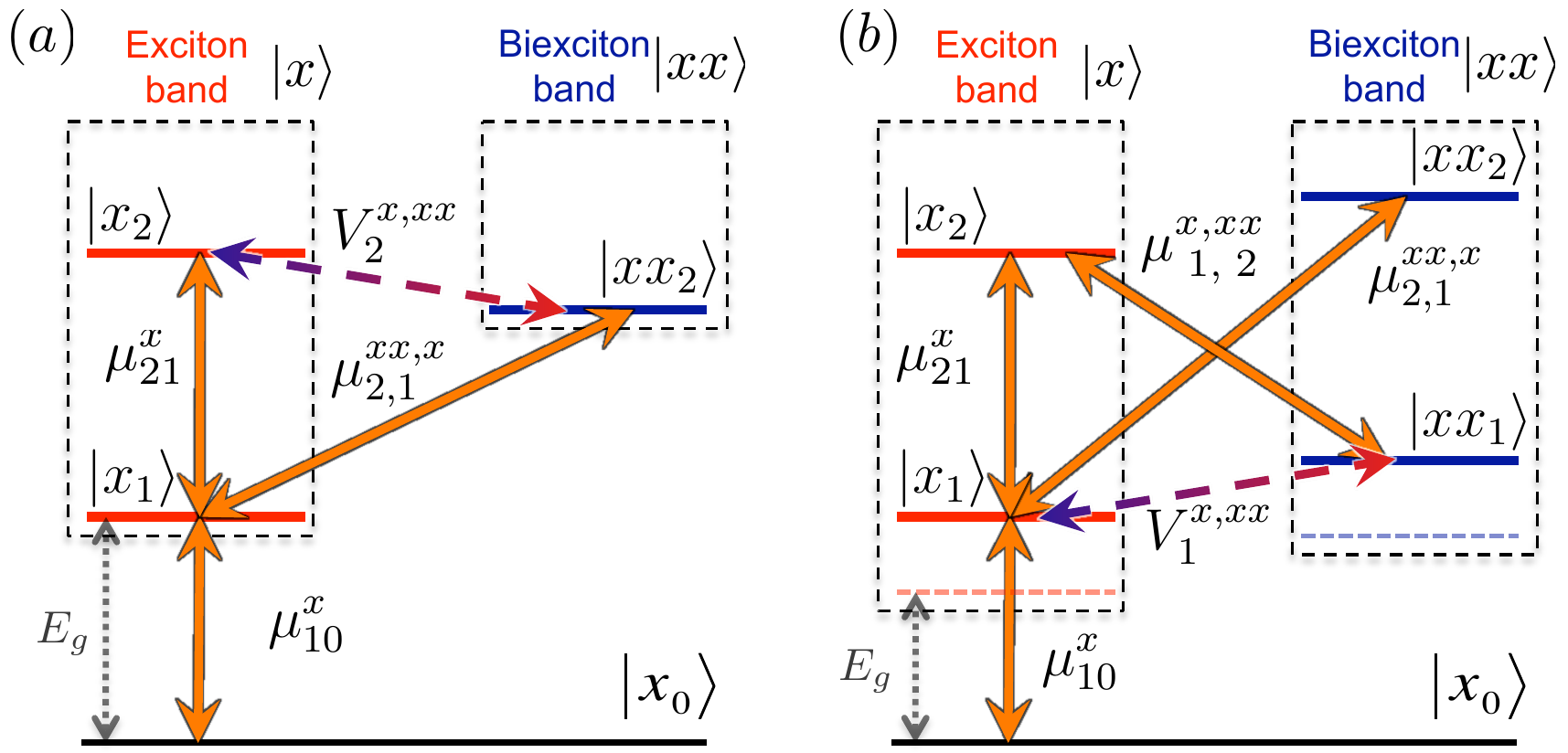}
\caption{Level diagrams for five-level systems reflecting different excitation regimes in NCs: 
		(a) $\hbar\omega_0\approx 2 E_g$ and (b) $\hbar\omega_0\gtrsim 2E_g$. Solid arrows show 
		optically allowed transitions and dashed arrows stay for the interband Coulomb interactions.
		Accordingly, the signal associated with panel~(a) has contribution from $S_1$, $S_2$, $S_3$, 
		and $S_4$ pathways shown in Fig.~\ref{fig-ld}. The signal associated with panel (b) is 
		contributed by $S_1$, $S_2$, $S_5$, and $S_6$ pathways shown in Fig.~\ref{fig-ld}.
		} 
\label{fig-fls}
\end{figure}

To identify unique signatures of the interband interactions in the 2D spectra, we need to reduce the number of cross-peaks by adopting a reduced level scheme shown in Fig.~\ref{fig-fls}~(b). Comparison of this model with $V^{x,xx}_2$ interaction added and generic multi-levle scheme (Fig.~\ref{fig-ld}) shows that the former one captures all the transition pathways present in the general case. As a result, double-quantum coherence signal from the five-level system also has eight non-vanishing components explicitly calculated in Appendix~\ref{app-fls}. Obtained expressions (Eq.~(\ref{Sj-fls})-(\ref{LmX}) and Table~\ref{tab-Sj-fls}) are used in our numerical calculations. The analysis of the signal can be further simplified by considering two excitation regimes, schematically shown in panels~(a) and (b) of Fig.~\ref{fig-fls}. These regimes are based on specific properties of exciton and biexciton level structure in semiconductor NCs. They also allow one to eliminate the contributions of $S_7$ and $S_8$ which include two interband propagation events (Fig.~\ref{fig-ld}) complicating the cross-peak patterns.       

We assume that considered semiconductor NCs possess inversion symmetry  leading to the parity conservation. We further assume that the lowest exciton state is optically active (``bright'') and, thus, has odd parity. Since, the lowest biexciton state is formed from the two lowest exciton states, it has even parity. The interband Coulomb interaction conserves parity and, therefore, does not couple exciton and biexciton states with different parities. Such an exciton and biexciton level structure is realized in lead-salt semiconductors. Therefore, without lack of generality, we adopt the effective mass model based on the ${\bf k}\cdot{\bf p}$-Hamiltonian for PbSe NCs.\cite{kang97} The diameter of the NCs is set to 5~nm. Since the calculations of the exciton and biexciton binding energies, as well as the interband Coulomb matrix elements are not directly accessible from the adopted model, we will consider these quantities as parameters. To explore 2D  signatures of the interband interactions, the interactions strengths will be varied in a broad range.

\subsection{Excitation Regime $\hbar\omega_0\approx E_g$.}
\label{fls-a}

First, we consider the case shown in panel~(a) of Fig.~\ref{fig-fls}, in which pulse central frequency is resonant with the lowest exciton state, $|x_1\rangle$, i.e., $\hbar\omega_0\approx E_g$. Under this condition, the double excitation energy corresponds to the Coulomb coupled even-parity lowest biexciton state, $|xx_2\rangle$ and even-parity exciton state $|x_2\rangle$. These sates are allowed for optical transitions from $|x_1\rangle$. According to the adopted effective mass model, we set $\hbar\omega^x_1=1.0$~eV, $\hbar\omega^x_2=2.2$~eV, and $\hbar\omega^{xx}_2=1.8$~eV with the binding energies included. Noteworthy, this model is similar to the so-called coherent superposition model introduced by Shabaev, Efros, and Nozik to study CM in PbSe nanocrystals.\cite{shabaev06} Since, the interband transition dipoles $\mu^x_{10}$ and $\mu^{xx,x}_{21}$ depend on the same valence-to-conduction band matrix elements, we set them equal. The intraband exciton transition dipole, $\mu^x_{21}$, vanishes in bulk semiconductors due to the quasimomentum conservation constraint. In NCs, this constraint is relaxed and the intraband transitions can acquire some oscillator strength. Therefore,  we consider three different cases in which $\mu^x_{21}=0$, $\mu^x_{21}=0.5\mu^{x}_{10}$, and $\mu^x_{21}=\mu^{x}_{10}$. In general, double-quantum coherence signal associated with so defined level scheme is a superposition of $S_1$, $S_2$, $S_3$, and $S_4$ components (Appendix~\ref{app-fls} and Fig.~\ref{fig-ld}).     

\begin{figure*}[t]
\centering
\includegraphics[width=4.5in,clip]{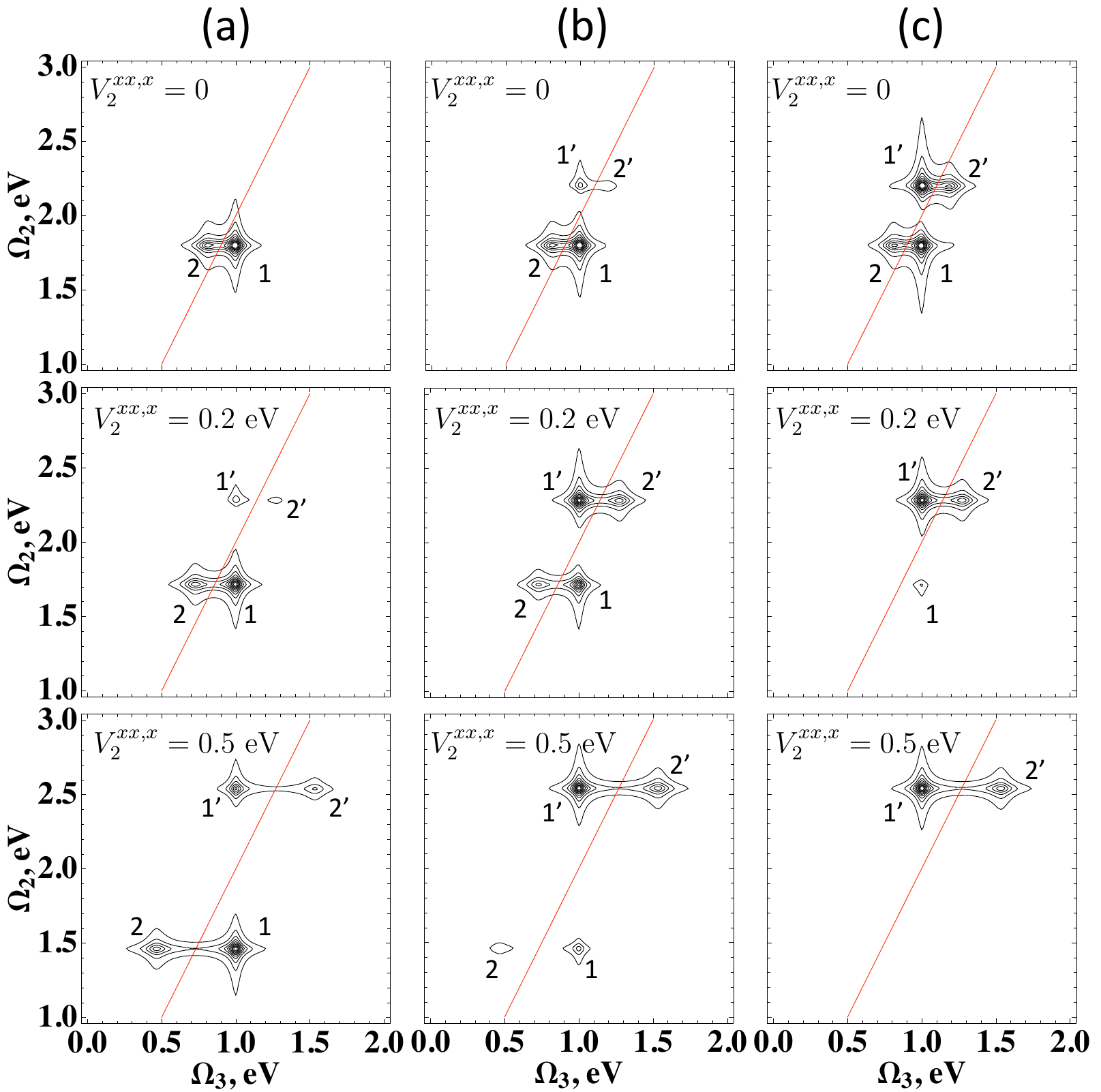}
\caption{The absolute value of double-quantum coherence signal for the excitation regime $\hbar\omega_0\approx E_g$ in NCs shown schematically in Fig.~\ref{fig-fls}~(a). The interband exciton dipole is set to (a) $\mu^x_{21}=0$, (b) $\mu^x_{21}=0.5\mu^{x}_{10}$, and (c) $\mu^x_{21}=\mu^{x}_{10}$. Within each column the interband Coulomb interaction increases from top to bottom taking the following values: $V^{x,xx}_2=0.0$, $V^{x,xx}_2=0.2$~eV and $V^{x,xx}_2=0.5$~eV. Red line stands for $\Omega_2=2\Omega_3$.}
\label{fig-2Da}
\end{figure*}

To start, we set $\mu^x_{21}=0$ reducing the total signal to the $S_2$ component which is associated with the pathway involving no interband propagation (Fig.~\ref{fig-ld}). The corresponding 2D spectra are shown in column~(a) of Fig.~(\ref{fig-2Da}). For $V^{x,xx}_{2}=0$ (upper panel), the 2D spectrum contains two $(\Omega_3,\Omega_2)$-cross-peaks. Namely, peak 1 at $(\omega^x_1,\omega^{xx}_2)$ and peak 2 at $(\omega^{xx,x}_{2,1},\omega^{xx}_2)$. The difference between the $\Omega_3$-resonances in this case provides the biexciton binding energy defined as $\omega^{xx}_2-2\omega^x_1$. The middle and lower panels in column~(a) show the same spectra for finite interband Coulomb coupling strength $V^{x,xx}_2=0.2$ eV and $V^{x,xx}_2=0.5$~eV, respectively.  Switching on the interaction leads to the renormalization of the peak 1 and 2 frequencies as $(\omega^x_1,\omega_{2-})$ and $(\omega_{2-}-\omega^x_1,\omega_{2-})$, respectively. In addition cross-peaks $1'$ at $(\omega^x_1,\omega_{2+})$ and $2'$ at $(\omega_{2+}-\omega^x_1,\omega_{2+})$ emerge. Here, $\omega_{2\pm}$ denotes  mixed (quasiparticle) energy of coupled $|x_2\rangle$ and $|xx_2\rangle$ states given by Eq.~(\ref{wpm}). Accordingly, the increase in the interband coupling leads to the increase in the energy splitting between the cross-peaks as observed in Fig.~\ref{fig-2Da}. 

Setting $\mu^x_{21}$ to finite values at $V^{x,xx}_{2}=0$ (upper panels in column~(b) and (c) of Fig.~\ref{fig-2Da}), adds $S_1$ contribution to the 2D signal which is solely associated with the intraband exciton transitions (Fig.~\ref{fig-ld}). This leads to the appearance of already seen cross-peaks $1'$ and $2'$, whose intensities increase as $\mu^x_{21}$ increases. Adding the interband Coulomb interaction (middle and lower panels in column~(b) and (c)) adds the contributions from the $S_3$ and $S_4$ signal components associated with the interband propagation (Fig.~\ref{fig-ld}). However, the resonances associated with these components overlap with already existing cross-peaks $1$, $2$, and $1'$, $2'$. 

The interference of all pathways contributing to the cross-peaks can lead to dramatic variation in their intensities. As observed in middle and lower panels of column~(b), the increase of the Coulomb coupling reduces the intensities of cross-peaks 1 and 2. For equal strength of inter- and intraband transition dipoles (column~(c)) and strong interband coupling (lower panel), cross-peaks $1$, $2$ vanish. This effect can be rationalized through the formation of symmetric, $|2+\rangle=1/\sqrt{2}(|x_2\rangle+|xx_2\rangle)$  and antisymmetric, $|2-\rangle=1/\sqrt{2}(|x_2\rangle-|xx_2\rangle)$, superpositions of the coupled states where the latter one is optically ``dark".      

Finally, we note that in the excitation regime $\hbar\omega_0\approx E_g$, the contributions of signal components $S_1$ through $S_4$ correspond to the cross-peak patterns that do not allow one to qualitatively distinguish between the effect of the interband Coulomb interactions associated with the components $S_3$ and $S_4$ (Fig.~\ref{fig-ld}) and the intraband dipole transitions accounted for  in $S_1$ (Fig.~(\ref{fig-ld})). This conclusion immediately follows from the comparison of the middle panel in column~(a) and the upper one in column~(b), which have the same cross-peak patterns. Therefore, considered excitation regime does not allow one to obtain a clear signatures of the interband Coulomb interactions. As we show next, the situation is quite different for the excitation regime shown in Fig.~\ref{fig-fls}~(b).   

\subsection{Excitation Regime $\hbar\omega_0\gtrsim 2E_g$.}
\label{fls-b}

According to the adopted effective mass model, the lowest odd-parity biexciton state in NCs is second-to-lowest biexciton state whose energy is slightly higher than $2E_g$. This biexciton is Coulomb coupled to an exciton state of the same parity. Since the latter exciton state is optically allowed, we turn pulse central frequency in resonance with this state. This leads to the excitation regime shown in Fig.~\ref{fig-fls}~(b). Following, the notations used here for the five-level system, we denote the coupled exciton and biexciton states as $|x_1\rangle$ and $|xx_1\rangle$, respectively. Solving the effective mass model for PbSe NCs\cite{kang97} as well as adding the binding energies, we set $\hbar\omega^x_1=2.2$~eV and $\hbar\omega^{xx}_1=2.5$~eV. We further chose a pair of double-excited even-parity exciton, $|x_2\rangle$ and biexciton, $|xx_2\rangle$, states with the energies $\hbar\omega^x_2=3.6$ eV and $\hbar\omega^{xx}_2=4.1$~eV, respectively. These states can be coupled through the interband Coulomb interactions. However, they belong to the high-energy region where these interactions are much weaker than $V^{x,xx}_1$ and therefore we set $V^{x,xx}_2=0$. Similar to Sec.~\ref{fls-a}, the interband transition dipoles $\mu^x_{10}$, $\mu^{xx,x}_{2,\;1}$, and $\mu^{xx,x}_{\;1,\;2}$ are set identical, and the intraband ones are varied as  $\mu^x_{12}=\mu^{xx}_{12}=0$ and  $\mu^x_{12}=\mu^{xx}_{12}=\mu^{x}_{10}$. In general, double-quantum coherence signal associated with such defined level scheme (Fig.~\ref{fig-fls}~{a}) is a superposition of $S_1$, $S_2$, $S_5$, and $S_6$ components (Appendix~\ref{app-fls} and Fig.~\ref{fig-ld}).

\begin{figure}[t]
\centering
\includegraphics[width=3.2in,clip]{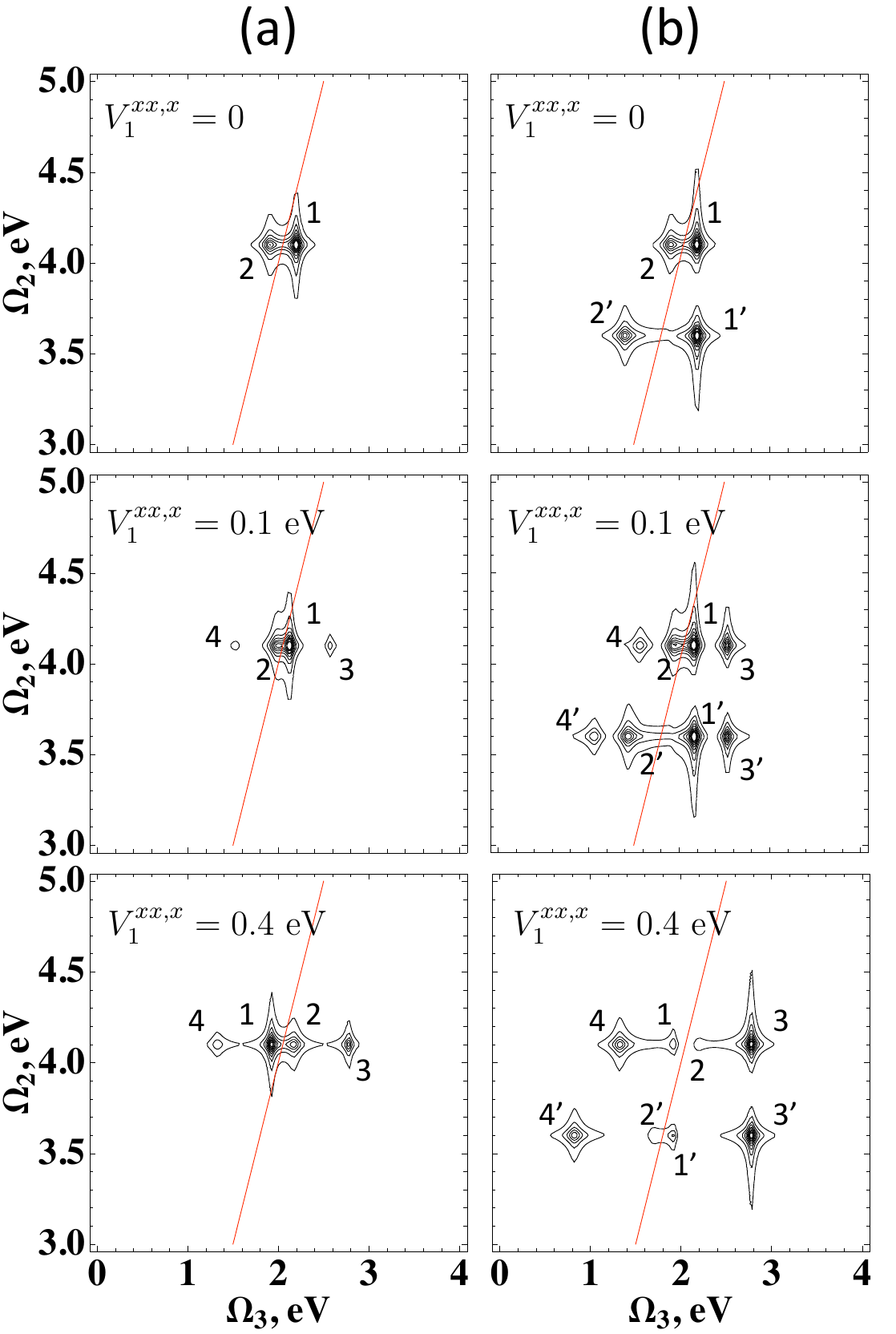}
\caption{The absolute value of double-quantum coherence signal for the excitation regime $\hbar\omega_0\gtrsim 2E_g$ in NCs shown schematically in Fig.~\ref{fig-fls}~(b). The interband transition dipoles are set to (a) $\mu^x_{21}=\mu^{xx}_{21}=0$, (b) $\mu^x_{21}=\mu^{xx}_{21}=\mu^{x}_{10}$. Within each column the interband Coulomb interaction increases from top to bottom taking the following values: $V^{x,xx}_2=0.0$, $V^{x,xx}_1=0.1$~eV and $V^{x,xx}_1=0.4$~eV. Red line stands for $\Omega_2=2\Omega_3$.}
\label{fig-2Db}
\end{figure}

Fig.~\ref{fig-2Db}~(a) shows double-quantum coherence spectra in the case when the intraband optical transitions are forbidden, i.e., $\mu^x_{21}=\mu^{xx}_{21}=0$. If interband Coulomb interaction is set to $V^{x,xx}_2=0$ (upper panel in column~(a)), the spectrum is solely due to $S_2$  component and contains cross-peaks 1 and 2 at $(\omega^x_1,\omega^{xx}_2)$ and $(\omega^{xx,x}_{2,1},\omega^{xx}_2)$, respectively. This is the same situation as we discussed in Sec.~\ref{fls-a}. In contrast to Sec.~\ref{fls-a}, turning on the interband interaction does not lead to the level splitting along $\Omega_2$, since $V_2^{x,xx}=0$. Instead, the interband interaction, $V^{x,xx}_1$, renormalizes 1,2 cross-peak frequencies as $(\omega_{1-},\omega^{xx}_2)$ and $(\omega^{xx}_{2}-\omega_{1-},\omega^{xx}_2)$ and leads to new features 3 and 4 emerging at the frequencies $(\omega_{1+},\omega^{xx}_2)$ and $(\omega^{xx}_{2}-\omega_{1+},\omega^{xx}_2)$, respectively.    

Switching on the intraband optical transitions adds a new set of resonances at $\Omega_2=\omega^x_2$ as can be seen in Fig.~\ref{fig-2Db}~(b). Specifically, the cross-peaks $1'$, $2'$, $3'$, and $4'$ emerge at $(\omega_{1-},\omega^{x}_2)$, $(\omega^{x}_{2}-\omega_{1-},\omega^{x}_2)$, $(\omega_{1+},\omega^{x}_2)$ and $(\omega^{x}_{2}-\omega_{1+},\omega^{x}_2)$, respectively. These resonances result from the contributions of $S_1$, $S_5$, and $S_6$ signal components. These components also provide additional contributions to cross-peaks $1-4$. As the Coulomb interaction rises from top to bottom of column~(b), the interference effect similar to that in Fig.~\ref{fig-2Da}~(b) and (c), affects the intensity of the cross-peaks.

Comparison of columns~(a) and (b) in Fig.~(\ref{fig-2Da}), shows that the adopted excitation regime allows one to distinguish between the effects of the interband Coulomb interaction and intraband optical transitions: If $V^{x,xx}_1=0$ and $\mu^x_{21}=\mu^{xx}_{21}=0$, one observes a pair of cross-peaks 1, 2 along $\Omega_3$. The peaks positions are symmetric with respect to the line $2\Omega_3-\Omega_2=0$ (red in Fig.~\ref{fig-2Da} and \ref{fig-2Db}), and the splitting provides the biexciton binding energy. For $V^{x,xx}_1=0$ and $\mu^x_{21}=\mu^{xx}_{21}\neq 0$, a new symmetric pair of cross-peaks $1'$ and $2'$ shows up  at $\Omega_2=\omega^x_2$. For $V^{xx}_1\neq 0$ and $\mu^x_{21}=\mu^{xx}_{21}=0$, two pairs of cross-peaks (i.e. 1, 2 and 3, 4) symmetric with respect to the line $2\Omega_3-\Omega_2=0$ are observed. In this case the energy splitting between symmetric peaks is due to the biexciton binding energy renormalized by the interband Coulomb coupling. Finally, for $V^{xx}_1\neq 0$ and $\mu^x_{21}=\mu^{xx}_{21}\neq 0$ all eight cross-peaks (i.e. $1-4$ and $1'-4'$) show up in the 2D spectrum. Note, that the symmetry of the cross-peak pairs with respect to $2\Omega_3-\Omega_2=0$ line can be used to determine the biexciton binding energies and the interband Coulomb interactions in more realistic situation, when large number of the resonances make their identification difficult. 

\section{Conclusions}
\label{conc}

Using previously developed exciton scattering models accounting for the Coulomb interactions between exciton and biexciton states, we have derived closed expressions for 2D double-quantum coherence signal. In general, the Liouville space pathways contributing to the signal account for the interband scattering processes and the effect of the inter- and intraband optical transitions. The interplay of these two effects is important to take into account considering spectroscopic probe of photoexcited carrier dynamics in semiconductor NCs. Presence of the intraband optical transitions and the interband scattering effect corresponds to the formation of new cross-peaks in the 2D spectra. To find clear spectroscopic signatures of each of the effects, we have considered two excitation regimes and found that the last one, i.e. $\hbar\omega_0\gtrsim 2E_g$ carries desired sensitivity. 

Although, the calculations performed here used effective mass model for PbSe NCs they can be considered rather model since unrealistic values for the biexciton binding energies and interband Coulomb couplings have been used. The analysis of realistic spectra would require more computational efforts and necessity to deal with large number of cross-peaks. However, the trends we have identified in the 2D spectra using our model system should be present and eventually sough in more realistic calculations and in the interpretation of experimental observations.

\acknowledgments
This work was supported by the Office of Basic Energy Sciences, US Department of Energy, and Los Alamos LDRD funds. We also acknowledge the support provided by CNLS. We wish to thank Shaul Mukamel pointing our attention to considered spectroscopic technique in context of the carrier multiplication problem.

\appendix
\section{The Exciton Scattering Formalism}
\label{app-esm}

In this Appendix, we provide a closed set of Eqs.~(\ref{Gf-Four})--(\ref{TF-inter}) allowing one to calculate the Hilbert space Green function (Eq.~(\ref{G-mtrx})) using the scattering matrix formalism. Derivation details can be found in Ref.~[\onlinecite{piryatinski10}].

We start with the Fourier representation of the time-domain Green function  
\begin{eqnarray}
    \label{Gf-Four}
    \hat G(t)= \int_{-\infty}^\infty\frac{d\omega}{2\pi}\hat G(\omega)\exp{\left(-i\omega t\right)},
\end{eqnarray}
in terms of the  frequency domain counterpart 
\begin{eqnarray}\label{GW-mtrx}
 \hat G(\omega) = \left( \begin{array}{cc}
					\hat G^{x}(\omega)    & \hat G^{x,xx}(\omega)\\
					\hat G^{xx,x}(\omega) & \hat G^{xx}(\omega)  \\
	\end{array} \right).
\end{eqnarray}
This quantity satisfies the following equation 
\begin{eqnarray}\label{GT-rep}
    \hat G(\omega)= \hat g(\omega) + \hat g(\omega)\hat T(\omega)\hat g(\omega),
\end{eqnarray}
where $\hat g(\omega)$ is the intraband free propagator
\begin{eqnarray}\label{g-mtrx}
 \hat g(\omega) = \left( \begin{array}{cc}
					\hat g^{x}(\omega)    & 0\\
					0 & \hat g^{xx}(\omega)  \\
	\end{array} \right),
\end{eqnarray}
and $\hat T(\omega)$ is the scattering operator
\begin{eqnarray}\label{T-mtrx}
 \hat T(\omega) = \left( \begin{array}{cc}
					\hat T^{x}(\omega)    & \hat T^{x,xx}(\omega)\\
					\hat T^{xx,x}(\omega) & \hat T^{xx}(\omega)  \\
	\end{array} \right),
\end{eqnarray}
containing exciton scattering, $\hat T^x$,  biexciton scattering, $\hat T^{xx}$, and the interband scattering, $T^{x,xx}$, components. 

For the projected Hamiltonian given by Eqs.~(\ref{Hs})--(\ref{HW}), the matrix elements of exciton ($\bar n = x$) and biexciton ($\bar n = xx$) free propagator have the following form
\begin{eqnarray}\label{g-wc}
    g^{\bar n}_{kl}(\omega) =\frac{i\delta_{kl}}{\omega-\tilde\omega^{\bar n}_k},
\end{eqnarray}
where the complex frequency, $\tilde\omega^{\bar n}_{k}=\omega^{\bar n}_k-i\gamma^{\bar n}_k$, contains the $k$-th frequency, $\omega^{\bar n}_k$, from the projected Hamiltonian (Eq.~(\ref{H0})), and the dephasing rate, $\gamma^{\bar n}_k$. The exciton ($\bar n=x$) and biexciton ($\bar n=xx$) scattering matrix elements satisfy a set of linear equations  
\begin{eqnarray}
    \label{TF-intra}
    \sum_{kl}\left[\delta_{ik}\delta_{lj}-i\sigma^{\bar n}_{ik}(\omega)g^{\bar n}_{kl}(\omega)\right]
    T_{lj}^{\bar n}(\omega)= i\sigma^{\bar n}_{ij}(\omega),
\end{eqnarray}
with the self-energy matrix elements defined as
\begin{eqnarray}
    \label{Selfe-eq}
    \sigma_{ij}^{\bar n}(\omega)= i\hbar^{-2}\sum_{kl} V_{i,k}^{\bar n,\bar m}
    g^{\bar m}_{kl}(\omega)V_{l,j}^{\bar m,\bar n},
\end{eqnarray}
where $\bar m=xx$ ($\bar m=x$) if $\bar n=x$ ($\bar n=xx$). Finally, the interband scattering matrix elements can be obtained from the following linear transformation: 
\begin{eqnarray}
  \label{TF-inter}
   T^{x,xx}_{a,l}(\omega)&=& -iV^{x,xx}_{a,\;\;l}
   -i\sum_{mn} V^{x,xx}_{a,\;\;m}g^{xx}_{mn}(\omega)T^{xx}_{nl}(\omega).\;\;\;\;\;\;
\end{eqnarray}

\section{2D Signal in Weak Coulomb Limit}
\label{app-wcl}

In this Appendix, we use weak Coulomb limit of the exciton scattering model\cite{piryatinski10} to derive the double-quantum coherence signal. Adopted limiting case assumes that the interband Coulomb coupling is weak compared to the energy difference between coupled states and/or their broadening. This ultimately leads to the leading contribution of the interband Born (single) scattering event.  
 
Retaining only leading Born contributions in Eqs.~(\ref{Gf-Four})--(\ref{TF-inter}), one finds that the diagonal blocks in Eq.~(\ref{G-mtrx}) are\cite{piryatinski10} 
\begin{eqnarray}
\label{GX}
    \hat G^{x}(t)&=& \sum_{a\geq 1}|x_a\rangle\langle x_a| e^{-i\tilde\omega^x_{a}t},
\\\label{GXX}
    \hat G^{xx}(t)&=& \sum_{k\geq 1}|xx_k\rangle\langle xx_k|e^{-i\tilde\omega^{xx}_{k}t}.
\end{eqnarray}
Here, the complex frequencies $\tilde\omega^x_{a} = \omega^x_{a}-i\gamma^{x}_{a}$ and $\tilde\omega^{xx}_{k} = \omega^{xx}_{k}-i\gamma^{xx}_{k}$ contain exciton, $\gamma^{x}_{a}$, and biexciton, $\gamma^{xx}_{k}$, dephasing rates, respectively. 

In the adopted limit, the leading off-diagonal blocks in Eq.~(\ref{G-mtrx}) become\cite{piryatinski10} 
\begin{eqnarray}
\label{GXXX}
    \hat G^{x,xx}(t)&=& \sum_{a\geq 1}\sum_{k\geq 1}|x_a\rangle 
    \Lambda^{x,xx}_{a,k}\langle xx_k|
\\\nonumber&\times&    
    (e^{-i\tilde\omega^x_at}-e^{-i\tilde\omega^{xx}_kt}),
\end{eqnarray}
where the interband transition amplitude is
\begin{eqnarray}\label{Lmbd}
    \Lambda^{x,xx}_{a,k}&=& \frac{V^{x,xx}_{a,k}}{\hbar\left(\tilde\omega^x_a-\tilde\omega^{xx}_k\right)}.
\end{eqnarray}

According to Eqs.~(\ref{mu-mtrx})--(\ref{S2D}), and the Green function representation given by Eqs.~(\ref{GXL})--(\ref{GXXXL}),  double-quantum coherence signal becomes a sum of six components
\begin{eqnarray}\label{Sall-wc}
 S(\Omega_3,\Omega_2) =\sum_{j=1}^6 S_{j}(\Omega_3,\Omega_2),
\end{eqnarray}
where each component is a limiting value of the general signal components given by Eq.~(\ref{Sj}) and can be identified with the corresponding pathways in Fig.~\ref{fig-ld}. Terms $S_7$ and $S_8$ are neglected, since they have second order in $V^{x,xx}$ contributions which are beyond the Born approximation. 

The first two signal components provide zero-order in $V^{x,xx}$ contributions  
\begin{eqnarray}\label{S2d-01}
 S_1(\Omega_3,\Omega_2) &=& \sum_{abc}\frac{\mu^x_{0c}\mu^x_{cb}\mu^x_{ba}\mu^x_{a0}}
 	{\Omega_2-\tilde\omega^x_{b}}
 \\\nonumber&\times&
 \left(\frac{1}	{\Omega_3-\tilde\omega^x_{c}}-\frac{1}{\Omega_3-\tilde\omega^x_{bc}}\right),
\end{eqnarray}
with $\tilde\omega^{x}_{bc}=\tilde\omega^{x}_{b}-\tilde\omega^{x}_{c}{}^*$, and
\begin{eqnarray}\label{S2d-02}
S_2(\Omega_3,\Omega_2) &=&\sum_{ac,n} \frac{\mu^x_{0c}\mu^{x,xx}_{cn}\mu^{xx,x}_{na}\mu^x_{a0}}
 	{\Omega_2-\tilde\omega^{xx}_{n}}
	\\\nonumber
  &\times&\left(\frac{1}{\Omega_3-\tilde\omega^x_{c}}
  -\frac{1}{\Omega_3-\tilde\omega^{xx,x}_{\;\;n,c}}\right),
\end{eqnarray}
with $\tilde\omega^{xx,x}_{n,\;c}=\tilde\omega^{xx}_{n}-\tilde\omega^{x}_{c}{}^*$. 

Next two terms, i.e., $S_j$ with $j=3,4$
\begin{eqnarray}
\label{S2d-11}
	S_j(\Omega_3,\Omega_2)&=&-\sum_{bc,n}\frac{F^j_{bc,n}}{\Omega_2-\tilde\omega^x_{b}}
\\\nonumber&\times&
			\left(\frac{1}{\Omega_3-\tilde\omega^x_{c}}-\frac{1}{\Omega_3-\tilde\omega^x_{bc}}\right)
	\\\nonumber&+&			
			\sum_{bc,n}\frac{F_{bc,n}}{\Omega_2-\tilde\omega^{xx}_{n}}
\\\nonumber&\times&
		\left(\frac{1}{\Omega_3-\tilde\omega^x_{c}}-\frac{1}{\Omega_3-\tilde\omega^{xx,x}_{\;\;n,c}}\right),
\end{eqnarray}
with 
\begin{eqnarray}
\label{F3}
	F^3_{abc,n}&=&\mu^x_{0c}\mu^{x,xx}_{cn}\Lambda^{xx,x}_{\;n,\;b}\mu^x_{ba}\mu^x_{a0}
\\\label{F4}
F^4_{abc,n}&=&\mu^x_{0c}\mu^{x}_{cb}\Lambda^{x,xx}_{b,\;n}\mu^{xx,x}_{\;n,\;a}\mu^x_{a0},
\end{eqnarray}
provide first order corrections to Eq.~(\ref{S2d-01}) and Eq.~(\ref{S2d-02}) Note that the resonances in Eq.~(\ref{S2d-11}) coincide with the resonances in Eq.~(\ref{S2d-01}) and Eq.~(\ref{S2d-02}).
  
The last two contributions to the signal read 
\begin{eqnarray}\label{S2d-13}
 	S_3(\Omega_3,\Omega_2) &=&\frac{1}{\hbar}\sum_{abc,m}
 			\mu^x_{0c}\Lambda^{x,xx}_{c,\;m}\mu^{xx,x}_{\;m,b}\mu^x_{ba}\mu^x_{a0}
	\\\nonumber &\times&  
 		\frac{\omega^{x,xx}_{c,\;m}}{\Omega_2-\tilde\omega^x_{b}}
			\left(\frac{1}{\Omega_3-\tilde\omega^x_{c})(\Omega_3-\tilde\omega^{xx}_{m})}
	\right.\\\nonumber &+&\left.	
			\frac{1}{(\Omega_3-\tilde\omega^x_{bc})(\Omega_3-\tilde\omega^{x,xx}_{b,\;m})}\right),
\end{eqnarray}
and
\begin{eqnarray}\label{S2d-14}
 	S^{(1)}_4(\Omega_3,\Omega_2) &=&\frac{1}{\hbar}\sum_{ac,nm}
 			\mu^x_{0c}\Lambda^{x,xx}_{\;c,\;m}\mu^{xx}_{mn}\mu^{xx,x}_{na}\mu^x_{a0}
	\\\nonumber &\times&  
 		\frac{\omega^{x,xx}_{c,\;m}}{\Omega_2-\tilde\omega^{xx}_{n}}
			\left(\frac{1}{(\Omega_3-\tilde\omega^x_{c})(\Omega_3-\tilde\omega^{xx}_{m})}
	\right.\\\nonumber &+&\left.	
			\frac{1}{(\Omega_3-\tilde\omega^{xx,x}_{\;n,\;c})(\Omega_3-\tilde\omega^{xx}_{nm})}\right),
\end{eqnarray}
These terms also provide first order corrections but contain new cross-peaks associated with the interband scattering. Specifically, these new peaks appear at $\Omega_3=\tilde\omega^{xx}_{m}$, $\Omega_3=\tilde\omega^{x,xx}_{b,\;m}$, and $\Omega_3=\tilde\omega^{xx}_{nm}$.

\section{2D Signal associated with Five-Level System}
\label{app-fls}

In this Appendix, we derive the expressions for double-quantum coherence signal associated with five-level system shown in Fig.~\ref{fig-fls}~(b). For completeness, we also add the interband Coulomb interaction $V^{x,xx}_2$ in the high-energy region.  

In this case, the non-interacting exciton and biexciton components of the projected Hamiltonian (Eqs.~(\ref{H0})--(\ref{HW})) reduce to the following form
\begin{eqnarray}\label{H0-fls}
 \hat{H}_0&=& \sum_{a=1,2}|x_a\rangle\hbar\omega^x_a\langle x_a|
    + \sum_{k=1,2}|xx_k\rangle\hbar\omega^{xx}_k\langle xx_k|.\;\;
 \end{eqnarray}
Here, the exciton energies are set to be $\hbar\omega^{x}_1\sim\omega_0$ and $\hbar\omega^{x}_2\sim 2\omega_0$. The difference between adjacent exciton and biexciton energies is assumed to be within the pulse spectral widths $\sigma\ll\omega_0$, i.e., $|\omega^x_p-\omega^{xx}_p|<\sigma$ for $p=1,2$. The adjacent exciton and biexciton states are coupled by the interband Coulomb interaction 
\begin{eqnarray}\label{HW-fls}
 \hat{V}_C &=& \sum_{c=1,2}|x_c\rangle V^{x,xx}_{c}\langle xx_c|+h.c..
\end{eqnarray}
Finally, the transition dipole operator for the adopted five-level system reads
\begin{eqnarray}\label{mu-fls}
  \hat\mu &=& |x_0\rangle \mu^{x}_{01}\langle x_1|+ |x_1\rangle \mu^{x}_{12}\langle x_2|
  \\\nonumber&+&
  |x_1\rangle \mu^{x,xx}_{1,\;2}\langle xx_2|+ |xx_1\rangle \mu^{xx,x}_{\;1,\;2}\langle x_2|
  \\\nonumber&+& 
  |xx_1\rangle \mu^{xx}_{12}\langle xx_2|+h.c..
 \end{eqnarray}

\begin{table}[t]
   \centering
   \begin{tabular}{m{1.5cm}|c|c|c|c} 
\hline\hline
Number of Interband Propagation Events &\;\;$S_j$\;\;  &\;$M_j$\;&\;$A_j$\;&\;$B_j$\;\\
\hline\vspace{0.25cm}
\multirow{2}{*}{\hspace{0.7cm}0}& $S_1$  & $|\mu^{x}_{01}|^2|\mu^{x}_{12}|^2$ &\;\;$\Lambda^{x}_{1\alpha}$\;\; &\;\;$\Lambda^x_{2\beta}$\;\;\\[0.25cm]
	 &$S_2$  & $|\mu^{x}_{01}|^2|\mu^{x,xx}_{\;1,\;2}|^2$ &\;\;$\Lambda^{x}_{1\alpha}$\;\; &\;\;$\Lambda^{xx}_{2\beta}$\;\;\\[0.25cm]
\hline\vspace{0.25cm}
\multirow{4}{*}{\hspace{0.7cm}1}& $S_3$   & $|\mu^{x}_{01}|^2\mu^{x,xx}_{1,2}\mu^{x}_{21}$ &\;\;$\Lambda^{x}_{1\alpha}$\;\; &\;\;$\Lambda^{xx,x}_{2\beta}$\;\;\\[0.25cm]
 	& $S_4$   & $|\mu^{x}_{01}|^2\mu^{x}_{21}\mu^{x,xx}_{2,\;1}$  &\;\;$\Lambda^{x}_{1\alpha}$\;\; &\;\;$\Lambda^{x,xx}_{2\beta}$\;\;\\[0.25cm]
 	& $S_5$   & $|\mu^{x}_{01}|^2\mu^{xx,x}_{\;1,\;2}\mu^{x}_{21}$ &\;\;$\Lambda^{x,xx}_{1\alpha}$\;\; &\;\;$\Lambda^{x}_{2\beta}$\;\;\\[0.25cm]
 	& $S_6$   & $|\mu^{x}_{01}|^2\mu^{xx}_{1,2}\mu^{xx,x}_{\;2,\;1}$ &\;\;$\Lambda^{x,xx}_{1\alpha}$\;\; &\;\;$\Lambda^{xx}_{2\beta}$\;\;\\[0.25cm]
\hline\vspace{0.25cm}
\multirow{2}{*}{\hspace{0.7cm}2}	& $S_7$   & $|\mu^{x}_{01}|^2|\mu^{x}_{12}|^2$ &\;\;$\Lambda^{x,xx}_{1\alpha}$\;\; &\;\;$\Lambda^{xx,x}_{2\beta}$\;\;\\[0.25cm]
 	& $S_8$   & $|\mu^{x}_{01}|^2|\mu^{x,xx}_{\;1,\;2}|^2$ &\;\;$\Lambda^{x,xx}_{1\alpha}$\;\; &\;\;$\Lambda^{x,xx}_{2\beta}$\;\;\\[0.25cm]
\hline\hline  
   \end{tabular}
   \caption{List of coefficients associated with the double-quantum coherence signal components given in Eq.~(\ref{Sj-fls}).
   }
   \label{tab-Sj-fls}
\end{table}

For so defined model, all eight double-quantum coherence signal components (Eq.~(\ref{Sj}) and Table~\ref{tab-Sj}), reduce to the following form
\begin{eqnarray}\label{Sj-fls}
 S_{j}(\Omega_3,\Omega_2) &=& M_j\sum_{\alpha,\beta=\pm}
 \frac{B_j}{\Omega_2-\tilde\omega_{2\beta}}
 \\\nonumber&\times&
 \left(\frac{A_j}{\Omega_3-\tilde\omega_{1\alpha}}
 		-\frac{A_j^*}{\Omega_3-\tilde\omega_{2\beta,1\alpha}}\right),
\end{eqnarray}
where $\tilde\omega_{2\beta,1\alpha}=\tilde\omega_{2\beta}-\tilde\omega_{1\alpha}^*$, and the quasiparticle energies are
\begin{eqnarray}
    \label{wpm}
    \omega_{c\pm}=\frac{\omega_c^{x}+\omega_c^{xx}}{2}\pm
    \sqrt{\left(\frac{\omega^{x}_c-\omega^{xx}_c}{2}\right)^2
    +\left(\frac{V^{x,xx}_c}{\hbar}\right)^2},
\end{eqnarray}
with $c=1,2$. The expressions for the coefficients $A_j$ and $B_j$ are listed in Table~\ref{tab-Sj-fls} and contain the following  transition amplitudes\cite{piryatinski10}
\begin{eqnarray}
\label{LmX}
    \Lambda^x_{c\pm}&=& \pm\frac{\left(\omega_{c\pm}-\omega_c^{xx}\right)}
                {\left(\omega_{c+}-\omega_{c-}\right)},
\\\label{LmXX}
    \Lambda^{xx}_{c\pm}&=& \pm\frac{\left(\omega_{c\pm}-\omega_c^{x}\right)}
                {\left(\omega_{c+}-\omega_{c-}\right)},    
\\\label{LmInt}
    \Lambda^{x,xx}_{c\pm}&=& \pm\frac{V^{x,xx}_c}
            {\hbar\left(\omega_{c+}-\omega_{c-}\right)},
\end{eqnarray}
with $c=1,2$.

To obtain the signal considered in Sec.~\ref{fls-a} (Fig.~\ref{fig-fls}~(a)), one has to set $V^{x,xx}_1=0$. This reduces contributions to the sum of $S_1$, $S_2$, $S_3$, and $S_4$ components. The signal considered in Sec.~\ref{fls-b} (Fig.~\ref{fig-fls}~(b))  can be obtained by setting $V^{x,xx}_2=0$. This results in the following non-vanishing contributions: $S_1$, $S_2$, $S_5$, and $S_6$.

\bibliographystyle{prsty}

\begin{thebibliography}{10}

\bibitem{haug09}
H. Haug and S.~W. Koch, {\em Quantum Theory of the Optical and Electronic
  Properties of Semiconductors} (World Scientific, London, 2009).

\bibitem{haug08}
H.~J.~W. Haug and A.-P. Jauho, {\em Quantum Kinetics in Transport and Optics of
  Semiconductors} (Springer-Verlag, Berlin, 2008).

\bibitem{axt98}
V.~M. Axt and S. Mukamel, Rev.~Mod.~Phys. {\bf 70},  145   (1998).

\bibitem{efros89}
A.~L. Efros and A.~V. Rodina, Solid~State~Commun. {\bf 72},  645   (1989).

\bibitem{banyai88}
L. Banyai, Y.~Z. Hu, M. Lindberg, and S.~W. Koch, Phys.~Rev.~B {\bf 38},  8142
   (1988).

\bibitem{banyai89}
L. Banyai, Phys.~Rev.~B {\bf 39},  8022   (1989).

\bibitem{takagahara89}
T. Takagahara, Phys.~Rev.~B {\bf 39},  10206   (1989).

\bibitem{hu90}
Y.~Z. Hu, M. Lindberg, and S.~W. Koch, Phys.~Rev.~B {\bf 42},  1713   (1990).

\bibitem{butov02}
L.~V. Butov, A.~C. Gossard, and D.~S. Chemla, Nature {\bf 418},  751   (2002).

\bibitem{piryatinski07a}
A. Piryatinski, S.~A. Ivanov, S. Tretiak, and V.~I. Klimov, Nano Lett. {\bf 7},
   108  (2007).

\bibitem{klimov07}
V. Klimov {\it et~al.}, Nature {\bf 447},  441  (2007).

\bibitem{landsberg03}
P.~T. Landsberg, {\em Recombination in Semiconductors} (Cambridge University
  Press, New York, 2003).

\bibitem{schaller04}
R.~D. Schaller and V.~I. Klimov, Phys.~Rev.~Lett. {\bf 92},  186601  (2004).

\bibitem{ellingson05}
R.~J. Ellingson {\it et~al.}, Nano Lett. {\bf 5},  865   (2005).

\bibitem{nair08}
G. Nair, S.~M. Geyer, L.-Y. Chang, and M.~G. Bawendi, Phys.~Rev.~B {\bf 78},
  125325  (2008).

\bibitem{mcguire08}
J.~A. McGuire {\it et~al.}, Acc.~Chem.~Res. {\bf 41},  1810  (2008).

\bibitem{pijpers09}
J.~J.~H. Pijpers {\it et~al.}, Nature~Phys. {\bf 5},  811  (2009).

\bibitem{kolodinski93}
S. Kolodinski, J. Werner, T. Wittchen, and H. Queisser, Appl.~Phys.~Lett. {\bf
  63},  2405  (1993).

\bibitem{nozik02}
A.~J. Nozik, Physica~E {\bf 14},  115   (2002).

\bibitem{hanna06}
M. Hanna and A. Nozik, J.~Appl.~Phys. {\bf 100},  074510  (2006).

\bibitem{klimov06}
V. Klimov, Appl.~Phys.~Lett. {\bf 89},  123118  (2006).

\bibitem{beard08a}
M.~C. Beard and R.~J. Ellingson, Laser~\&~Photon.~Rev. {\bf 2},  377  (2008).

\bibitem{nozik08}
A.~J. Nozik, Chem.~Phys.~Lett. {\bf 457},  3  (2008).

\bibitem{luther08}
J.~M. Luther {\it et~al.}, Nano Lett. {\bf 8},  3488  (2008).

\bibitem{schaller05}
R.~D. Schaller, V.~M. Agranovich, and V.~I. Klimov, Nature~Phys. {\bf 1},  189
   (2005).

\bibitem{rupasov07}
V.~I. Rupasov and V.~I. Klimov, Phys.~Rev.~B {\bf 76},  125321  (2007).

\bibitem{shabaev06}
A. Shabaev, A.~L. Efros, and A.~J. Nozik, Nano Lett. {\bf 6},  2856   (2006).

\bibitem{witzel10}
W.~M. Witzel {\it et~al.}, Phys.~Rev.~Lett. {\bf 105},  137401  (2010).

\bibitem{piryatinski10}
A. Piryatinski and K.~A. Velizhanin, J.~Chem.~Phys. {\bf 133},  084508  (2010).

\bibitem{mukamel_book}
S. Mukamel, {\em Principles of Nonlinear Optical Spectroscopy} (Oxford
  University Press, Oxford, 1995).

\bibitem{mukamel00}
S. Mukamel, Annu.~Rev.~Phys.~Chem. {\bf 51},  691  (2000).

\bibitem{mukamel04}
S. Mukamel and D. Abramavicius, Chem.~Rev. {\bf 104},  2073   (2004).

\bibitem{li06}
X. Li, T. Zhang, C.~N. Borca, and S.~T. Cundiff, Phys.~Rev.~Lett. {\bf 96},
  057406  (2006).

\bibitem{mukamel07}
S. Mukamel, R. Oszwaldowski, and D. Abramavicius, Phys.~Rev.~B {\bf 75},
  245305  (2007).

\bibitem{oszwaldowski08}
R. Oszwaldowski, D. Abramavicius, and S. Mukamel, J.~Phys.~Cond.~Matt. {\bf
  20},  045206  (2008).

\bibitem{ernst90}
R.~R. Ernst, G. Bodenhausen, and A. Wokaun, {\em Principles of Nuclear Magnetic
  Resonance in One and Two Dimensions} (Clarendon Press, Oxford, 1990).

\bibitem{mukamel07a}
S. Mukamel, R. Oszwaldowski, and L. Yang, J.~Chem.~Phys. {\bf 127},  221105
  (2007).

\bibitem{abramavicius08}
D. Abramavicius, D. Voronine, and S. Mukamel, Proc.~Natl.~Acad.~Sci.~USA {\bf
  105},  8525  (2008).

\bibitem{kim09}
J. Kim, S. Mukamel, and G.~D. Scholes, Acc.~Chem.~Res. {\bf 42},  1375  (2009).

\bibitem{yang08}
L. Yang and S. Mukamel, Phys.~Rev.~Lett. {\bf 100},  057402  (2008).

\bibitem{yang08a}
L. Yang and S. Mukamel, Phys. Rev. B {\bf 77},  075335  (2008).

\bibitem{stone09}
K.~W. Stone {\it et~al.}, Science {\bf 324},  1169   (2009).

\bibitem{stone09a}
K.~W. Stone {\it et~al.}, Acc.~Chem.~Res. {\bf 42},  1452   (2009).

\bibitem{karaiskaj10}
D. Karaiskaj {\it et~al.}, Phys. Rev. Lett. {\bf 104},  117401  (2010).

\bibitem{kang97}
I. Kang and F.~W. Wise, J.~Opt.~Soc.~Am.~B {\bf 14},  1632   (1997).

\end{thebibliography}

\end{document}